\documentclass{PoS}
\usepackage{subfigure}

\title{High-resolution studies of the twin jet in NGC\,1052}

\ShortTitle{The twin jet in NGC\,1052}

\author{\speaker{E.~Ros$^a$}\,, M.~Kadler$^{b,c,d,e}$\,, Y.~Y.~Kovalev$^{a,f,}$\thanks{Alexander von Humboldt Fellow}\,, H.~D.~Aller$^{g}$\,, M.~F.~Aller$^{g}$\,, S.~Kaufmann$^{h}$\\
        \llap{$^a$}Max-Planck-Institut f\"ur Radioastronomie, Auf dem H\"ugel 69, D-53121 Bonn, Germany\\
        \llap{$^b$}Dr.\ Remeis-Sternwarte Bamberg, Universit\"at Erlangen-N\"urnberg, Sternwartstrasse 7, D-96049 Bamberg, Germany\\
        \llap{$^c$}Erlangen Centre for Astroparticle Physics, Erwin-Rommel-Str.\ 1, D-91058 Erlangen, Germany\\
        \llap{$^d$}CRESST/NASA Goddard Space Flight Center, Greenbelt, MD 20771, USA\\
        \llap{$^e$}Universities Space Research Association, 10211 Wincopin Circle, Suite 500 Columbia, MD 21044, USA\\
        \llap{$^f$}Astro Space Center of Lebedev Physical Institute, Profsoyuznaya 84/32, 117997 Moscow, Russia\\
        \llap{$^g$}Department of Astronomy, University of Michigan, 817 Denison Building, Ann Arbor, MI 48109-1042, USA\\
        \llap{$^h$}Landessternwarte Heidelberg, Universit\"at Heidelberg, K\"onigstuhl 12, D-69117 Heidelberg, Germany\\
        E-mail: \email{ros@mpifr.de}, 
        \email{matthias.kadler@sternwarte.uni-erlangen.de},
        \email{ykovalev@mpifr.de},
        \email{haller@umich.edu},
        \email{mfa@umich.edu}
        \email{skaufmann@lsw.uni-heidelberg.de}
        }

\abstract{
The radio loud galaxy NGC\,1052 is being studied in an
intensive multi-band campaign including X-ray brigthness monitoring
and spectroscopic observations, single-dish radio brightness monitoring
at centimetre wavelengths, and a high-frequency 
very-long-baseline interferometry monitoring program.  
Here we present a progress report on our studies from this program.
The final goal of 
our observations is to relate the findings from the 
high-resolution radio images with the observed variations
in the X-ray regime, to address the accretion processes and
their relationship with the radio jet activity.
}

\FullConference{The 9th European VLBI Network Symposium on The role of VLBI in the Golden Age for Radio Astronomy and EVN Users Meeting\\
		 September 23-26, 2008\\
		 Bologna, Italy}

\begin{document}

\section{Introduction}
The radio source NGC\,1052 (at a distance of 21.6\,Mpc) offers
one of the best opportunities to study the nature of active
galactic nuclei (AGN).  This source is radio loud and can be classified
as a Type 2 AGN.  It shows a twin-jet system in
east-west direction at VLA and VLBI
scales, oriented close to the plane of the 
sky \cite{vermeulen03,cooper07}.  
Multi-wavelength VLBI observations
have shown evidence for an obscuring torus at the central region
of the source, hiding the inner part of the western jet at
longer wavelengths \cite{kellermann99,kameno01,kadler04}.  
This source has been monitored at the 2\,cm~Survey/MOJAVE programmes
since 1995.  A kinematic analysis of the sub-parsec scale morphology
of the source shows apparent speeds of 0.26\,c in the jet
and the counter-jet \cite{vermeulen03,ros08}, with ejection of
new features (or components) every 3 to 6 months.

\section{The multi-band campaign}

Evidence in support of
an accretion-ejection event was provided by observations around epoch
2001.0 \cite{kadler05}, where variations in the relativistic
broad iron K$\alpha$ profiles were seen before and
after a VLBI component ejection.  This motivated us to start
a multi-mission campaign of observations of this source in mid 2005.
Previous results have been reported in \cite{ros07,ros08}.
Here we show further progress in this program.

\subsection{X-ray observations}
\begin{figure}
	\centering
		\includegraphics[clip,width=0.98\textwidth]{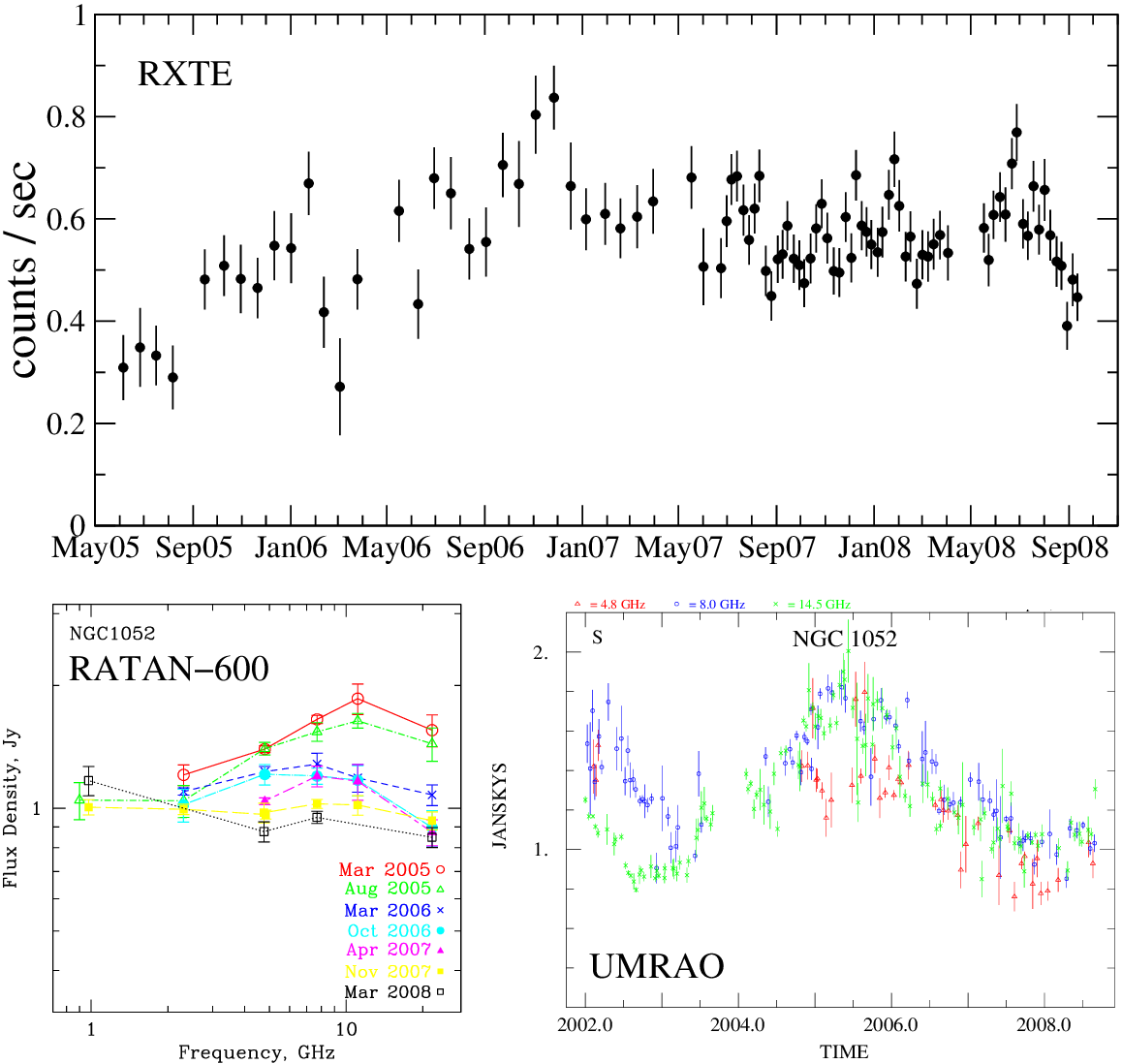}
	\caption{Brightness variability in NGC\,1052.  \textit{Top:} X-ray light curve 
taken by \emph{RXTE} since 2005.  A rising phase is shown in late 2005 and late 2006, followed by a variable, non-flaring state since early 2007, and a dip in August 2008.  This dip, if the source is similar in behaviour to 3C\,120 (see \cite{marscher06}), could herald a new VLBI component.
\textit{Bottom, left:}
RATAN-600 single-dish radio spectra of NGC\,1052, measured during our observation campaign.  The flux density axis shows values in logarithmic scale between 0.316\,Jy and 3.16\,Jy.
\textit{Bottom, right:}
UMRAO single-dish radio light curves.
}
	\label{fig:fluxes}
\end{figure}

Our campaign includes:
\begin{itemize}
\item \emph{RXTE} observations once every three weeks until 2007.4 and weekly since then: preliminary results are shown at the top panel in Fig.~\ref{fig:fluxes}
\item \emph{XMM-Newton} deep spectroscopy in 2006.3 (complementing earlier measurements in 2002.62), and further observations approved for 2008.7 and 2009.2
\item \emph{Swift} observations with XRT imaging in 2007.05, 2007.59, 2008.74, 2008.81, 2008.82, 2008.83, and 2008.92, and BAT continuous hard X-ray (14--194\,keV) monitoring
\item \emph{Suzaku} spectroscopical observations on 2007.54: these results are presented in \cite{brenemann09}, showing a narrow core and a broader component of Fe-K$\alpha$ emission robustly detected at 6.4\,keV, together with a soft, thermal emission component below 1\,keV.
\end{itemize}

\subsection{Radio observations}
\begin{figure}
	\centering
		\includegraphics[clip,width=0.6\textwidth]{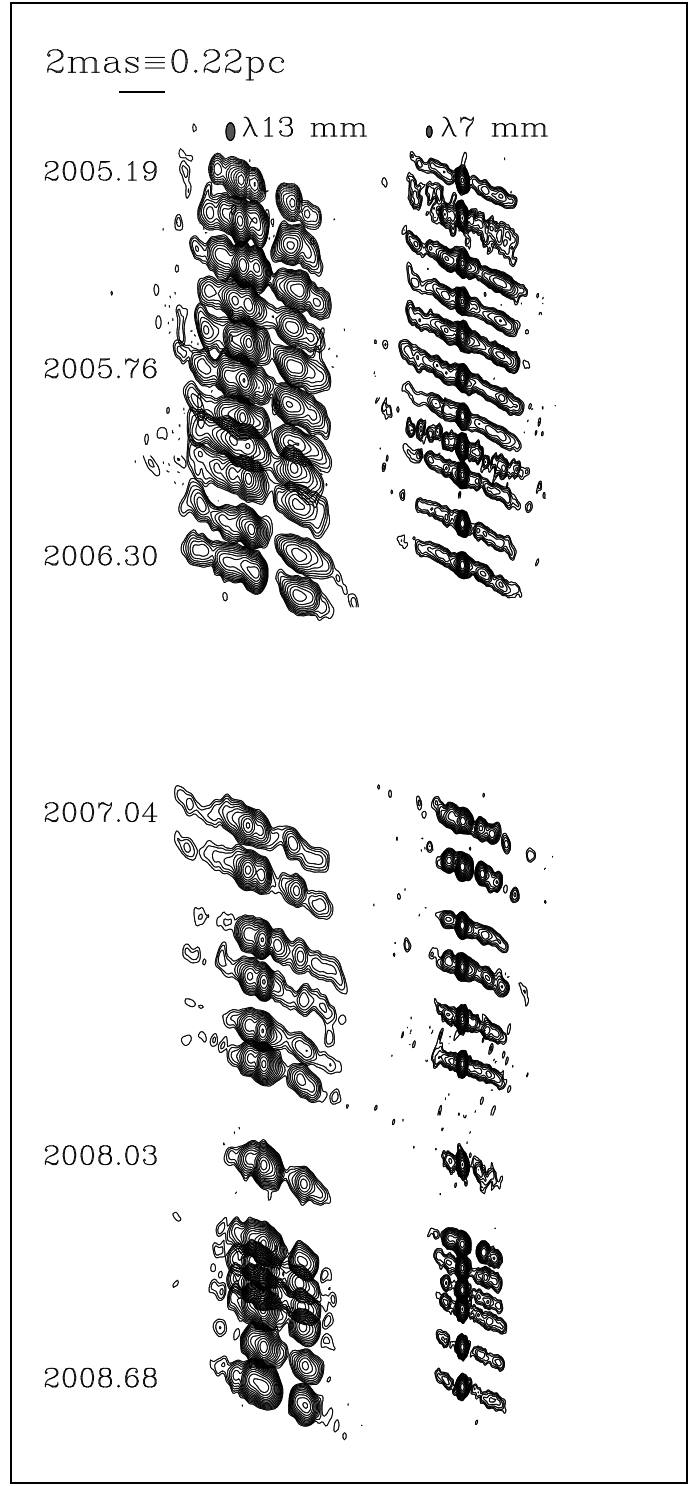}
		\caption{Preliminary VLBA images of NGC\,1052.  The alignment is arbitrary, to the emission gap at the $\lambda$1.3\,cm and to the brightness peak at $\lambda$0.7\,cm.}
		\label{fig:vlba}
\end{figure}

Our observations include single-dish and interferometric measurements as follow:
\begin{itemize}
\item Very Long Baseline Array (VLBA) $\lambda$2\,cm imaging observations in the framework of the 2\,cm~Survey/MOJAVE programmes \cite{kellermann04,lister05,lister09}, separated by several months (those observations are being performed since 1995)
\item VLBA $\lambda$1.3\,cm and $\lambda$0.7\,cm imaging observations separated by six weeks since 2005.19: preliminary images of these observations are shown in Fig.~\ref{fig:vlba}
\item Effelsberg single-dish flux density observations at seven wavelengths from $\lambda$21\,cm to $\lambda$1\,cm
\item RATAN-600 single-dish flux density observations at six wavelengths from $\lambda$30\,cm to $\lambda$1.3\,cm every four months (see \cite{kovalev99} for details): spectra measured at different epochs during our monitoring programme are shown at the left panel in Fig.~\ref{fig:fluxes}, where changes are visible at high frequencies, and no variations are observed at the lower ones, where the inner part of the jet is self-absorbed
\item University of Michigan Radio Astronomy Observatory (UMRAO) single-dish flux density observations at $\lambda$6\,cm, $\lambda$4\,cm and $\lambda$2\,cm--light curves at these frequencies is shown in Fig.~\ref{fig:fluxes}.
\end{itemize}

\section{Discussion}
Our campaign is providing radio light curves showing a quiet radio 
phase without bright flares since 2005.  
The UMRAO light curve shows the beginning of a rising 
phase in mid 2008, which gives hints of a new flare in the 
radio source.  
From the VLBA $\lambda$1.3\,cm and $\lambda$7\,cm radio images, 
two new features appear at the base of both jets, beginning to 
be visible after the first months of 2006.  
The X-ray curve shows dramatic drops in the X-ray light curve 
in February 2006, December 2006, and August 2008.  
A direct relationship between the appearance of new features and
the dips in the X-ray emission has still to be studied in detail.
Further progress in this campaign will include a detailed 
kinematic analysis of the observed radio features, after astrometric 
registration of those images, and the connection of these radio 
variations with the putative changes in the X-ray spectra.

\begin{footnotesize}
\paragraph*{Acknowledgements}
The Very Long Baseline Array is operated by the USA National Radio Astronomy Observatory, which is a facility of the USA National Science Foundation operated under cooperative agreement by Associated Universities, Inc.
\end{footnotesize}

\end{document}